# Concluding remarks: *Faraday Discussion* on astrochemistry at high resolution

T. J. Millar  *

Fifty years on from the first detailed chemical kinetic modelling of astronomical sources, I provide some introductory comments on the history of astrochemistry, summarise some personal views on the topics covered in this discussion meeting, and conclude with some thoughts on its future development. I have left out the jokes.

## 1 The past

Before 1970 only seven molecules were observed in interstellar space. CH, $CH^+$ and CN were detected in the period 1937–1941 through their optical absorption lines seen against the background of bright stars. The next molecular detections did not occur until the advent of radio astronomy and the development of receiver technology. Maser emission from OH was detected in 1963 and in the following six years ammonia ($NH_3$), water ($H_2O$) – again a maser – and formaldehyde ($H_2CO$), were identified. In their 1969 discovery paper of the latter molecule, Snyder *et al.*[1] noted that '$H_2CO$ ⋯ the first organic molecule ever detected in the interstellar medium ⋯ indicates that the process of chemical evolution may be much more complex than previously assumed'. With the current identification of close to 300 molecules, including the fullerenes $C_{60}$ and $C_{70}$, Snyder *et al.*[1] could not have imagined the range of species contained in interstellar and circumstellar environments.

The earliest attempt to describe the gas phase chemical processes that might operate in interstellar clouds was published by Bates and Spitzer[2] in an effort to understand the CH and $CH^+$ observations. With the exception of ion–neutral reactions, they identified most of the important chemical processes likely to operate in the diffuse environment in which these species were found. Indeed, finding that their calculated abundances did not reproduce the observations, they even considered the possibility that the thermal desorption of $CH_4$ from grains might be their source. Of course, one should recognise that a major reason for model failures was their lack of knowledge of the physical properties of the diffuse clouds in which these molecules were found.

*Astrophysics Research Centre, Queen's University Belfast, University Road, Belfast BT7 1NN, UK. E-mail: tom. millar@qub.ac.uk*

The first detailed numerical model of chemical kinetics in interstellar clouds was published by Herbst and Klemperer[3] and is rightly regarded as a classic in our field. The authors included many of the processes at work and worried about some that still concern us today – the elemental abundance of oxygen, the carbon-to-oxygen (C/O) elemental abundance ratio, the role of grains as a source and sink of interstellar molecules, and the extent to which a global chemical kinetic model can be applied to individual molecular clouds. Although some arguments concerning time scales are somewhat dated, this is, I would argue, the foundational paper for our subject and one which led directly to the plethora of experimental, theoretical, gas–grain modelling approaches to astrochemistry presented at this meeting.

Starting around 1985, spectral line surveys of high-mass star-formation systems began to detect in a large number of objects, a plethora of interstellar complex organic molecules (iCOMs), very loosely defined as containing at least six atoms including one element in addition to carbon and hydrogen. Many of the sources were rather hot and dense by interstellar standards, ∼100–300 K, $n(H_2) \sim 10^6$–$10^9$ cm$^{-3}$. These species, typified by their smallest member $CH_3OH$, are often saturated and tend to have larger abundances, often by several orders of magnitude, in these sources compared to those in cold (10 K) clouds. Observations of their D-substituted analogues, however, showed a surprising result – molecular D/H abundance ratios that are much larger, by factors of $10^3$–$10^5$, than the cosmic D/H ratio of a few $10^{-5}$. Indeed, the largest enhancement I am aware of occurs for triply-deuterated ammonia, $ND_3$, which has an abundance relative to $NH_3$ of about $10^{-3}$, some $10^{11}$ times larger than the statistical D/H ratio. Such large fractionation can only occur at very low temperatures and the accepted paradigm is that iCOMs and their deuterated versions are formed in molecular ices on cold dust grains before being desorbed into hot gas through the effect of stellar heating or shock waves. Such observations led to a number of models in which H atom collisions with a cold ice surface was assumed to convert atoms such as N, C, O and S to $NH_3$, $CH_4$, $H_2O$ and $H_2S$ in a water-rich ice mantle. Subsequent heating due to the presence of a newly-formed hot, massive star liberated large abundances of these simple hydrogenated molecules to the gas phase where subsequent gas-phase chemistry produced the iCOMs.[4–6] The lack of any detailed description of the actual chemistry on these icy grains and the cry of the modellers that 'all' abundance anomalies could be explained by surface chemistry led Malcolm Walmsley to note that grain chemistry 'was the last refuge of the scoundrel'.

## 2 The present

As indicated earlier, there are now 300 or so molecules detected in space with more than 75 containing 8 or more atoms (see Fig. 1).

In the last five years there has been a substantial increase in the number of detections of complex hydrocarbons mainly driven through dedicated spectral line surveys of the cold, dark Taurus Molecular Cloud 1 (TMC-1). McGuire and colleagues have concentrated on using sophisticated analysis of rotational emission lines as part of their GOTHAM survey to identify a number of cyclic hydrocarbons including benzonitrile,[7] 1- and 2-cyanocyclopentadiene, $C_5H_5CN$,[8,9] indene,[10] and 2-cyanoindene.[11] Cernicharo and co-workers, using the very sensitive NANOCOSMOS receiver, have detected many more hydrocarbons in their QUIJOTE survey of TMC-1 over the past few years, including several 5- and 6-membered ring species and branched chains. Interesting species include 1- and 2-ethynyl cyclopentadiene (c-$C_5H_5$CCH), ethynyl benzene ($C_6H_5$CCH),[12] and ethynyl cyclopropenylidene (c-$C_3$HCCH).[13]

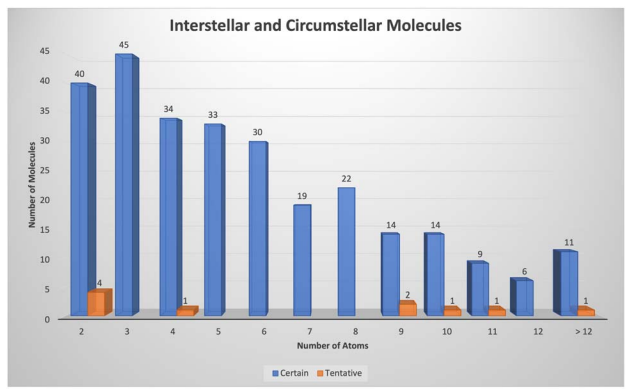

Fig. 1 Number of molecules detected in interstellar and circumstellar environments as a function of the number of atoms.

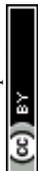

This discussion meeting was convened at a very appropriate time – and with a very timely agenda – given the recent development of very sensitive facilities such as the JWST (James Webb Space Telescope) with both higher sensitivity and spectral resolution than any other IR space observatory; as well as high spatial resolution instruments such as the ALMA (Atacama Large Millimeter/submillimeter Array). The early results from the JWST are breath-taking with an impact that affects every discipline in astronomy. The ALMA, which has been in operation for a decade, also continues to offer new views of the molecular universe and many of the observational papers at this meeting reflected the current dominant positions of these two facilities.

The meeting was arranged around four main themes: (1) observational astrochemistry; (2) laboratory gas-phase studies; (3) experimental studies of dust and ices; (4) computational astrochemistry. A number of poster papers were also on display during the meeting. My comments on the presentations are not meant to be comprehensive but reflect a number of personal factors including a lack of time to assimilate everything in the papers, presentations, extensive discussions and posters, a subjective bias toward certain topics, and a degree of ignorance in some areas. Nevertheless, I do commend the proceedings to the reader; they contain a wealth of information on both the rude health and the underlying concerns – the known unknowns – of our subject and its importance in answering fundamental questions in astrophysics, including the formation of stars and planetary systems. In particular, Cecilia Ceccarelli's Spiers Memorial Lecture (https://doi.org/10.1039/d3fd00106g), is a masterly introduction to the field, outlining recent successes, identifying complexities yet to be addressed in a comprehensive manner, and pointing to the future. For my part, the most important message of the meeting is that the foundation of astrochemistry is spectroscopy. Without the decades of advances in laboratory spectroscopy, and especially in astronomical instrumentation, our exploration of chemical processes in space would not be 'much more complex' than those considered 50 years ago.

### 2.1 Observational astrochemistry

Five papers discussed recent results from the JWST and ALMA. Van Dishoeck (https://doi.org/10.1039/d3fd00010a) and Kamp (https://doi.org/10.1039/d3fd00013c) presented a selection of spectacular results from the JWST mid-infrared (MIRI) spectral line surveys of protostars (JOYS) and of T Tauri and very low mass star (mostly brown dwarf) disks (MINDS). One clear advantage of the JWST observations over those from the ALMA, is that the former provides information on the abundances and excitation of non-polar molecules. The data presented were from a selection of the survey results but both papers show that there is a diversity within and between source types. Van Dishoeck's spectra show evidence of amorphous and crystalline silicates, ice bands, together with $H_2$ pure rotational and other molecular emission lines, most notably $H_2O$ and $CO_2$. The disk around the very low-mass star J160532 shows perhaps the most surprising spectrum: line rich in strong hydrocarbon emissions from $C_2H_2$, $C_4H_2$ and $C_6H_6$. Indeed in some sources the hydrocarbons are so abundant that they form a 'quasi-continuum'. This source also contains very intense $^{13}C^{12}CH_2$ emission. The analyses presented here indicate the presence of hot ($\sim$500 K) acetylene with $N(C_2H_2) \sim 10^{21}$ cm$^{-2}$, an enormous value that indicates that either efficient 'bottom-up' combustion chemistry is occurring or that carbonaceous grains are being chemically or thermally destroyed in a 'top-down' chemistry.

Kamp derived similar results from fitting a combination of slab and 2D thermochemical disk models to her observed spectra. She noted that line lists for $^{13}C^{12}CH_2$ were incomplete, preventing a full analysis of the spectra and of the (likely large) $^{13}C/^{12}C$ ratio. Again differences within and between source types were evident. All T Tauri disks contain $H_2O$ whereas it is only tentatively detected in the low mass disks. She also stressed the need to go beyond the slab retrieval models which treat only one molecule at a time and assume constant density and temperature. In addition, she noted that the underlying molecular line data and partition functions needed for such analyses are often missing. Her comparison between column densities derived with both slab and thermochemical models show that the latter typically predict values a factor of ten larger than the former. Clearly a more accurate approach is needed but again it was stressed that databases currently lack fundamental information – Einstein $A$-coefficients, partition functions at high $T$ and $P$ – as well as spectroscopy to provide hot band and isotopologue line lists for hydrocarbons. Finally, it is worth re-iterating that these are the early results from the JWST; we can expect a host of new species and new insights into the chemistry and physics of these planet-forming systems in the coming years. As always with new facilities, it will be a challenge for both laboratory and theoretical approaches to keep abreast of new observational developments.

Three papers presented high angular resolution ALMA observations on spatial scales ranging from the Solar System to $10^5$ times larger. On the largest scale, Huang's observations (https://doi.org/10.1039/d3fd00007a) used molecular tracers of shock waves to probe the similarities and differences between molecular gas in two galaxies – NGC 253, a starburst galaxy, and NGC 1068, an active galactic nucleus harbouring a supermassive black hole which generates high X-ray and cosmic-ray fluxes. Detailed observations were interpreted by models of shock chemistry finding evidence for the presence of both slow shocks, *via* observations of HNCO, and fast shocks, *via* the detection of large abundances of SiO which arise from the sputtering of silicate dust grains. Wilkins and Blake (https://doi.org/10.1039/d3fd00003f) presented a pilot spectral line search in the 1.2 mm window toward 11 giant molecular clouds (GMCs) at a distance of 4–8 kpc from the Galactic Centre, while Bianchi (https://doi.org/10.1039/d3fd00018d) investigated molecular line emission in SVS13-A, proto-binary system, in particular from the streamers connecting the binary system to their larger scale molecular envelope. This source has a rich array of molecular lines generated by iCOMs in the hot gas in the innermost region. Her study of HDO and $SO_2$ found a large, warm (95–160 K), column density of the former, indicating that it has been released from dust grain mantles, most likely by shock processes generated in the infalling gas. Astrochemistry and the astrophysics that can be derived from it are likely to be dominated by the ALMA and the JWST for the next decade or two – an exciting prospect.

### 2.2 Gas-phase laboratory studies

One striking aspect, at least to me, of this discussion meeting was that it is probably the first time that there were no papers on ion–neutral reactions, which have historically been the dominant mechanism for molecular synthesis studied in astrochemistry. In the main part, this dominance was due to the fact that both reactants and products could be detected readily whereas for neutral–neutral reactions it was more common to follow only the decay of one reactant with no measured information on products or branching ratios. Experimental gas-phase kinetics of neutral–neutral systems were covered in the papers by Suits on the reaction of vibrationally excited CN with butadiene (https://doi.org/10.1039/d3fd00029j), by Douglas on the low temperature behaviour of $NH_2$ with acetaldehyde, $CH_3CHO$ (https://doi.org/10.1039/d3fd00046j), and by Balucani (https://doi.org/10.1039/d3fd00057e) on the reaction of $N(^2D)$ with benzene. All three papers contain sophisticated, state-of-the-art experimental and theoretical techniques to detect products; the use of high-level quantum calculations to determine potential energy surfaces and the determination of total rate coefficients and branching ratios. The impetus for each study was to explain the formation of a specific interstellar molecule: Suits as a source of cyanobutadiene ($C_4H_5CN$), in TMC-1;[14] Douglas as a source of acetamide ($CH_3CONH_2$) in the interstellar medium;[15] and Balucani as a possible source of cyclopentadienyl (c-$C_5H_5$) in Titan. All required a significant investment of time and people. It is therefore rather disappointing for the authors, and for the wider astrochemical community, that none of the reported studies has had a significant impact on astrochemical models. I return to this point in Section 3.

The remainder of the session was devoted to spectroscopy and the study of transient species. This included papers by Brünken (https://doi.org/10.1039/d3fd00015j) on the fragmentation products of cyclic ions; Gupta who used a cryogenic ion trap at 4 K to record the rotational spectrum of c-$C_3H_2D$

(https://doi.org/10.1039/d3fd00068k); and Jacovella (https://doi.org/10.1039/d3fd00016h) who measured the rotational spectrum of the cross-branched molecule norbornadiene ($C_8H_7$), and its cyano-derivatives, which have much larger electric dipole moments, as a pre-cursor to an (unsuccessful) astronomical search. Puzzarini (https://doi.org/10.1039/d3fd00052d) presented a strategy for determining the rotational spectra of transient molecules using quantum calculations to predict properties and thereby guide laboratory searches and line identification.

### 2.3 Experimental studies of dust and ices

The third session, which concentrated on chemical reactions in and on dust grains, was marked by an animated and informed discussion session as perhaps befits a subject that is arguably the most important and the most uncertain aspect of understanding astrochemistry in molecular clouds and protoplanetary disks. Many astronomers could be forgiven for taking the view – to paraphrase the English essayist Charles Lamb's (1775–1834) comment about space and time:

'Nothing puzzles me more than molecules and ices yet nothing bothers me less as I never think about them'.

Nevertheless it is the job of the astrochemist to both worry and puzzle over the known and unknown unknowns of ice chemistry.

Four major questions on the cold (~10 K) ice chemistry that is thought to produce iCOMs, continue to exercise the community: what process removes these species from grains at temperatures of ~6–10 K to the cold gas phase in which they are detected?; do these processes return whole or fragmented species?; what are the binding energies of surface iCOMs since these determine the efficiency of thermal desorption processes?; and, given the fact that experiments and theory are limited to relatively few species, can we generalise results from the few to the many? A number of answers to these questions have been suggested in the literature and some were covered here.

Brown (https://doi.org/10.1039/d3fd00024a) discussed experiments which showed that infrared radiation could desorb CO from CO–water ice mixtures, noting that CO didn't desorb when irradiated with 4.6 μm photons but did so when the irradiation was at the wavelengths of the $H_2O$ vibrational modes at 2.9 and 12 μm. If a single photon process and if the efficiency is high, this could be an important desorption mechanism in dark clouds which are much more transparent to IR radiation than to UV photons, which have long been thought to drive photodesorption. Brown identified two mechanisms: a fast process of indirect resonant desorption in which an excited $H_2O$ molecule removes a neighbouring CO molecule, and a slow, photon-induced process in which energy is accumulated in the heat bath of water ice. It should be noted that the water ice layers are thin enough that the IR photons also heat the cold, metal substrate and that this may also contribute to desorption. Although poorly defined, Brown suggested yields of CO molecules removed per incident photon to be on the order of a few $10^{-5}$.

Like Brown, Bertin (https://doi.org/10.1039/d3fd00004d) studied the photodesorption of ices, in his case with vacuum UV (VUV) photons and ices composed of either pure formic acid (HCOOH), or methyl formate ($HCOOCH_3$), or with these species embedded in CO or $H_2O$ ices.

In addition to determining desorption yields, around $10^{-5}$ per incident VUV photon, another key issue is whether molecules are returned intact or as fragments. Bertin found similar results to previous experiments on methanol-containing ice, namely, that species detected in the gas phase are similar to the gas-phase photodesorption products of the molecule. Were this to be a universal rule it would be very convenient for modellers, but it is not – recent experiments on acetonitrile, $CH_3CN$, show that it photodesorbed intact.[16]

Gudipati (https://doi.org/10.1039/d3fd00048f) discussed an alternative desorption mechanism, one driven by the conversion of water ice from the amorphous to crystalline form. His experiments doped water ice with the volatiles CO, $CO_2$ and $O_2$ at 10 K and traced the outgassing species *via* mass spectrometry as the ice was heated to about 200 K. He showed that their abundance in the ice reduce by about a factor of 100 due to outgassing when the phase change occurs.

Finally, Sameera (https://doi.org/10.1039/d3fd00033h) discussed the important issues of binding energies and reaction barriers on ices. It has been known for some time, and certainly since the pioneering experimental work of Collings *et al.*,[17] that the binding energies of a molecule in water ice depends on the specific site in which it sits. More recently, Tinacci *et al.*[18] used theoretical calculations of $NH_3$ on a model cluster containing 100 $H_2O$ molecules to show that binding energies followed a double peaked profile with that at higher energies well fit by experimental TPD measurements. Rather surprisingly, however, there was a second distribution with much lower binding energies comparable to those of CO.

Sameera concentrated on $CH_3OH$ in water ice using a theoretical model to consider the reactions:

$$CH_3OH + OH \rightarrow CH_2OH + H_2O \quad (1)$$

$$CH_3OH + OH \rightarrow CH_3O + H_2O \quad (2)$$

His quantum mechanical calculations show that his model ice cluster has a number of binding sites and that the nature of these has a significant effect on the binding energies, on the computed reaction barriers, and on the detailed pathways that the reactants follow.

Many of the key issues raised here, including the effects of binding energies on grain chemistry, are discussed at length in Ceccarelli's Spiers Memorial Lecture (https://doi.org/10.1039/d3fd00106g).

### 2.4 Computational astrochemistry

This section contains a few papers which showed the complexity of computational approaches to chemical kinetics in the gas phase. I note that the focus of the meeting on astrochemistry meant that the complex issue of combining hydrodynamics with kinetics was not covered.

Garrod and Herbst (https://doi.org/10.1039/d3fd00014a) presented a very detailed time-dependent model of iCOM formation in molecular clouds based on their identification that several detected and candidate interstellar molecules containing the amine group, $-NH_2$, have proton affinities (PA) that are greater than that of $NH_3$, 853.6 kJ $mol^{-1}$. Because of its relatively large abundance in molecular clouds and its large PA, astrochemical models tend to include proton transfer reactions between ammonia and molecular ions as an important step in forming the neutral molecule. An example is:

$$H_2COH^+ + NH_3 \rightarrow NH_4^+ + H_2CO \quad (3)$$

The advantage of this proton transfer reaction is that it preserves the underlying neutral molecule, whereas the alternative loss of molecular ions, dissociative recombination with electrons, can destroy the molecule in a large fraction of cases. In the case of protonated formaldehyde, reaction with electrons has five product channels with that leading to formaldehyde occurring only around 30% of the time.[19] For protonated methanol, the situation is much worse with only 5% of recombinations re-cycling the neutral.[20] As a result, stable neutral molecules with PAs greater than that of ammonia are destroyed faster than those with PAs less than ammonia. Garrod's paper shows that this destruction process has a particular affect on the abundances of many amine species, including methylamine, ethylamine, urea and glycine. The latter molecule, yet to be identified in interstellar clouds, is a cornerstone in the argument that the precursors of life on Earth may be widely dispersed in the Galaxy – Garrod's results suggest that it may be more difficult to detect glycine than we currently envisage. One should note, however, that for reactions in which the difference between PAs is large, most importantly for reactions involving $H_3^+$, the excess reaction energy can lead to dissociative products rather than proton transfer, although not in every case.[21]

Viti (https://doi.org/10.1039/d3fd00008g) and Van de Sande (https://doi.org/10.1039/d3fd00039g) discussed novel ways of approaching the gas–grain modelling of molecular cloud and circumstellar chemistry, respectively. The former presented an alternative approach of using machine learning to model some of the uncertainties in complex chemical kinetic networks while the latter used a sophisticated physical model to discuss the effects of clumpy structures in the expanding envelopes around late-type stars. Viti's paper used both Bayesian and machine learning approaches to identify both those binding energies with the largest impact on calculated abundances and those species for which determining binding energies are a priority. Van de Sande's paper includes the words 'disentangling' and 'when adding complexity' in its title – a bold claim. As a co-author, I recuse myself at this point and leave it as an exercise for the reader as to whether the content lives up to its title!

Bromley (https://doi.org/10.1039/d3fd00055a) addressed the important topic of dust nucleation in the context of arguments – in essence that the dust destruction time scale is shorter than the dust formation time scale in evolved stars and supernovae – that a significant fraction of silicate dust must nucleate in the low density interstellar medium. Bromley concentrated on the creation of nanosilicates, combining quantum calculations on the lowest energy structures of small seed particles with their infrared absorption characteristics. He showed that the JWST could be sensitive enough to detect such particles if they composed 5–10% of the silicate grain population.

Finally, Madhusudhan (https://doi.org/10.1039/d3fd00075c) described the atmospheric physics and chemistry on a Hycean world, that is a sub-Neptunian exoplanet whose surface is covered by water oceans and with a $H_2$-rich atmosphere. Although no such planet has yet been detected, several candidate systems have been identified orbiting M dwarf stars. In addition to evolution time scales of several Gyr, Madhusudhan's paper showed one of the important differences between exoplanetary chemistry and astrochemistry, namely that the degree of physical complexity included in the former is of a much higher order than in the latter. In particular, the atmospheric models, internal planetary structures and delivery of exogenous materials are all informed by detailed observations and models of planet Earth and other solar-system bodies. Of course major uncertainties remain but he showed that the three essential ingredients for life – an energy source, liquid water and bio-essential elements – are all present in such worlds and that the JWST should be able to detect specific atmospheric molecules that are characteristic of Hycean worlds, an exciting prospect.

## 3  The future

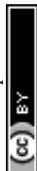

The development of astrochemistry over the past 50 years has been astounding, driven forward by the innovation of thousands of scientists, engineers and technicians whose ingenuity has led to the availability of facilities such as the ALMA and JWST. With these, we are able to probe astrochemistry from galactic size scales at high redshift, to au scales in protoplanetary disks and to even smaller scales in exoplanetary atmospheres. Coupled with these advances in technology, we have seen concomitant developments in chemical physics. Experimental and theoretical spectroscopy has underpinned the identification and analysis of many of the thousands of interstellar lines that have been detected by new facilities and instruments. Such work will continue to be important, particularly for the identification of metal-containing molecules, small reactive radicals, and large iCOMs and hydrocarbons, isotopologues and excited states.

Advances in chemical kinetics have also been dramatic over the past 50 years. Many important low-temperature experimental studies are now routinely supplemented by theoretical calculations of potential energy surfaces and reaction rate coefficients and product branching ratios. Such research, built on advanced computational techniques and computing power, is now routine and, although expensive and time-consuming, it can answer important questions in astrochemistry.

In Section 2, I noted that in recent years, observational searches coupled with high resolution spectroscopy have identified more than 100 iCOMs and complex hydrocarbon species in interstellar clouds. Their observed abundances have led to innovative chemical models including both gas-phase and surface reactions. For example, Garrod's basic model, as described here, has over 22 000 reactions among 1385 species of which roughly half are in the gas phase with the rest equally divided between surface ice and bulk ice. Yet for many species, and as we heard in the discussions, abundance calculations rest on 'thin ice' (pardon the pun).

Let me use, as an example, some results from my own, as yet unpublished work,[22] discovered while updating the UMIST Database for Astrochemistry chemical network, Rate12.[23] Propene ($CH_3CHCH_2$), is a key intermediary in the synthesis of large hydrocarbons. It was detected in TMC-1 with a relatively large fractional abundance, $4 \times 10^{-9}$, by Marcelino *et al.*[24] and was thought to form *via* two sequential radiative association reactions:

$$C_3H_3^+ + H_2 \rightarrow C_3H_5^+ + h\nu \quad (4)$$

$$C_3H_5^+ + H_2 \rightarrow C_3H_7^+ + h\nu \quad (5)$$

followed by dissociative recombination of $C_3H_7^+$ with electrons. The Rate12 database included this synthesis and showed that the calculated abundance agreed reasonably well with that observed. Subsequently, however, Lin et al.[25] showed that both radiative association pathways had barriers in their entrance channels and could not operate in cold clouds such as TMC-1. The exclusion of these two reactions leads to a huge drop in the calculated abundance of propene as shown in the Rate22 curve in Fig. 2.

This fall in its calculated abundance by about 10 orders of magnitude leads to a 'propene catastrophe' in the models since it is known experimentally to be a reactive molecule even at very low temperatures. For example, it has fast measured rate coefficients in reactions with CH (15 K),[26] CN (23 K),[27] $C_2$ (77 K),[28] $C_2H$ (15 K)[29] and $C_4H$ (39 K),[30] where the value in parentheses refers to the lowest laboratory temperature at which the rate coefficient has been measured. The products of these reactions include many of the hydrocarbons and their cyano-derivatives that have been detected in TMC-1, including c-$C_5H_6$, $CH_3C_4H$, $H_2CCCHCCH$, $CH_2CHCHCH_2$, $CH_3C_6H$, $CH_3CHCHCN$, $CH_2C(CH_3)CN$ and $CH_2CHCH_2CN$. Their calculated abundances are now more than $10^4$ times less than observed – and the misery goes on, since these species are also involved in the synthesis of other complex molecules including c-$C_5H_5CN$, c-$C_5H_4CCH_2$, c-$C_5H_5CCH$ and $H_2CCCHC_3N$. Identifying efficient, low-temperature formation pathways to propene is one of the most important tasks for astrochemistry today. Such 'single-point failures' also exist for other species in the networks. Indeed it is likely that somewhere in the region of 70–100 detected molecules are not yet included in even the most complex networks because their synthesis – in the gas phase or in ices – is unknown.

To conclude, let me mention four generic areas that we will need to cultivate in the coming years if we are to ensure that the field of astrochemistry continues both its remarkable growth over the past 10–20 years and its relevance to astrophysics far and near. The first is **synergy**. To me, one of the central messages of this discussion meeting – funded by the Royal Society of Chemistry – is that research in core theoretical and experimental physical chemistry disciplines, including spectroscopy and gas-phase kinetics, plays a central role in astrochemistry. Collaborations between observers and modellers within the astronomy community were evident in the papers describing JWST and ALMA results and noticeably increase their impact.

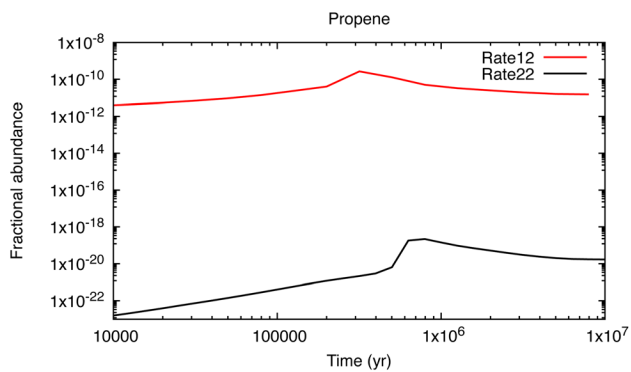

Fig. 2 The fractional abundance of propene relative to $H_2$ as a function of time for a model with properties similar to TMC-1 ($T = 10$ K, $n(H_2) = 10^4$ cm$^{-3}$).

Synergy across disciplines was also evident in and central to, many of the presentations and posters in this meeting and will increase in importance. For example, it is clear that many of the assumptions that underlie the grain-surface chemical kinetic models are too simplistic. Close co-operation between experimental and theoretical approaches on topics such as binding energies, surface and bulk diffusion, photodesorption yields, reaction products and desorption mechanisms, will be needed to ensure that theoretical models of interstellar chemistry become more predictive rather than reactive in nature. In a similar fashion, the astrochemical community needs to better identify critical reactions and processes for the chemical physics community to study.

The second is **strategy**. These synergistic links between chemistry and astronomy are often the result of individual actions, that is, where one researcher makes a measurement or calculation and reaches out to someone in the other field. There is an element of chance about these and indeed such interactions, while useful to a certain degree, may not have as much impact as one would like. We therefore need to find ways in which we can support research in a much more strategic manner. In the Netherlands, the Dutch Astrochemistry Network has been funded since 2010 and has supported many cross-disciplinary projects and, just as importantly, people. It not only brings together expertise in chemistry and astronomy but it also allows for the exchange of PhD students between institutions.

Such an approach is problematic in the UK where chemistry and astrophysics are funded by two separate research councils and where, in addition, many of the EU's cross-disciplinary programmes are ever more difficult to access. In the UK the joint RSC/RAS body, the Astrophysical Chemistry Group, provides a forum – but no funding – through which interested parties can interact.

There are major issues facing many of us in astrochemistry – can the astronomical community help shield experimental groups, often expensive to run, from the growing commodification of modern university operations, where the bottom line is, too often, profit? This is not a one-way support of course. Chemistry in its broadest sense is universal and as we move further into the era of the ALMA and the JWST, to more detailed explorations of exoplanet atmospheres and the planets, moons and minor bodies of our solar system, astronomers will need the chemical disciplines to help interpret and understand observational results.

The third area is **succession**. In my view, the paper by Herbst and Klemperer,[3] published 50 years ago, represents the birth of astrochemistry as a scientific discipline. Many of the scientists involved in the development of the foundations of the subject in its first 20 years, have retired or are close to retirement. They/we need to be active in planning their/our replacements. This means getting involved in departmental/university politics to ensure that the best people are recruited to, and promoted within, our discipline.

The final area is **students**. Our view of the molecular universe has improved enormously in the past 50 years. The development of high-resolution facilities, such as discussed here, provides us with an amazing panoply of images and spectra, coupled with a recognition of astrochemistry as central to answering many fundamental questions in astronomy, from the formation and evolution of galaxies to the formation of stars and planets. It was very noticeable at this meeting that the field has a wonderful array of talented, smart people who already possess a mature, informed, view of their research. The quality and impact of presentations and posters by PhD students and postdoctoral researchers was impressive as was their engagement in the discussion sessions. We need to ensure that, collectively, we harness those skills and encourage, excite, and enthuse our early career scientists. Where we can, let us support their intellectual growth, their exchange among institutions and across national boundaries, and let us find ways to employ them in full-time positions.

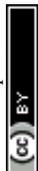

## Conflicts of interest

There are no conflicts of interest to declare.

## Acknowledgements

I am grateful to the Organising Committee for the invitation to present these closing remarks and to the Royal Society of Chemistry for the financial support that enabled me to do so in person. I am also grateful to the Leverhulme Trust for the award of an Emeritus Fellowship.

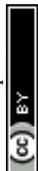